\documentclass[aip,apl,numerical,reprint,superscriptaddress]{revtex4-1}

\usepackage{graphicx}
\usepackage{latexsym}
\usepackage{amsmath}
\usepackage{amssymb}
\usepackage{amsfonts}
\usepackage{color}

\begin{document}
\bibliographystyle{aipsamp}

\title{Electromagnetic proximity effect in planar superconductor-ferromagnet structures}

\author{S. Mironov}
\affiliation{Institute for Physics of Microstructures, Russian Academy of Sciences,
603950 Nizhny Novgorod, GSP-105, Russia}
\author{A. S. Mel'nikov}
\affiliation{Institute for Physics of Microstructures, Russian Academy of Sciences,
603950 Nizhny Novgorod, GSP-105, Russia}
\affiliation{Lobachevsky State University of Nizhny Novgorod, 23 Gagarina, 603950
Nizhny Novgorod, Russia}
\author{A. Buzdin}
\affiliation{University Bordeaux, LOMA UMR-CNRS 5798, F-33405 Talence Cedex, France}
\affiliation{Department of Materials Science and Metallurgy, University of
Cambridge, CB3 0FS, Cambridge, United Kingdom}

\begin{abstract}
The spread of the Cooper pairs into the ferromagnet in proximity
coupled superconductor - ferromagnet (SF) structures is shown to cause a
strong inverse electromagnetic phenomenon, namely, the long-range transfer of
the magnetic field from the ferromagnet to the superconductor. Contrary to the
previously investigated inverse proximity effect resulting from the spin
polarization of superconducting surface layer, the characteristic length of
the above inverse electrodynamic effect is of the order of the London
penetration depth, which usually much larger than the superconducting
coherence length. The corresponding spontaneous currents appear even in the
absence of the stray field of the ferromagnet and are generated by the
vector-potential of magnetization near the S/F interface and they should be taken into account at the design of the nanoscale S/F devices. Similarly to the well-known Aharonov-Bohm effect, the discussed phenomenon can be viewed as a
manifestation of the role of vector potential in quantum physics.
\end{abstract}

\maketitle

The proximity phenomena in condensed matter physics are known to include the interface effects which are
usually associated with the exchange of electrons between the contacting materials. 
This exchange is responsible
for the mutual transfer of different particle-related qualities through the interface such 
as superconducting correlations, spin ordering etc. \cite{Buzdin_RMP, Nat1, Nat2}
The goal of the present work is to show that this spread of particle-related qualities in some cases should be supplemented
by the long-range spread of the electromagnetic fields. As an example, we demonstrate that
such electromagnetic proximity effect can strongly affect the physics of 
 superconductor - ferromagnet (SF) systems which are widely discussed as building blocks of superconducting spintronics \cite{Linder, Eschrig}.

\begin{figure*}[hbt]
\includegraphics[width=1\textwidth]{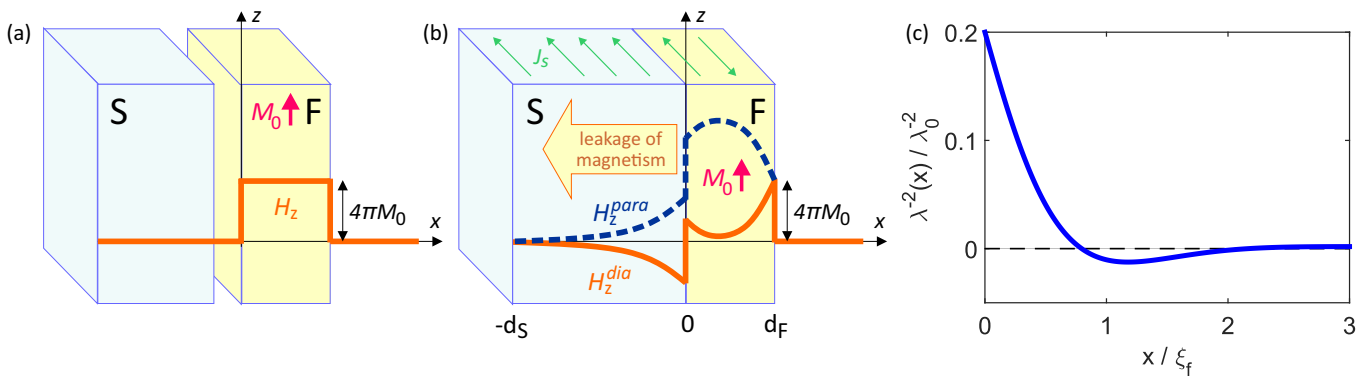}\caption{The sketch of the
superconductor/ferromagnet bilayer. (a) When the layers are separated from each other the magnetic field exists only inside the F layer. (b) In contrast, when the layers are put in contact the magnetization inside the ferromagnet becomes the source of the long-range magnetic field in the superconductor. In both panels the orange solid (blue dashed) curves schematically represent the
profile of the magnetic field when the total current inside the F-layer is diamagnetic (paramagnetic). (c) The spatial profile of the screening parameter $\lambda^{-2}(x)$ inside the ferromagnetic layer in the dirty limit. We take  $\sigma_s=10\sigma_f$.}%
\label{Fig_System}%
\end{figure*}

To elucidate our key observation we can consider an exemplary bilayer system
consisting of a superconducting (S) film placed in contact with a
ferromagnetic (F) layer with the magnetic moment parallel to the layer plane
(see Fig.~\ref{Fig_System}). Considering the S and F subsystems to be isolated we get a perfect
example of complete separation of the regions with a nonzero concentration of
Cooper pairs and magnetic field. The latter is completely trapped inside the
ferromagnet. As we allow the electron transfer between the subsystems the
Cooper pairs immediately penetrate the ferromagnet inducing the pair electron
correlations there. This process which is usually referred as a standard
proximity effect leads to the series of fascinating transport phenomena
\cite{Buzdin_RMP, Bergeret_RMP}. The inverse proximity effect
namely the transfer of the magnetic moment from the ferromagnetic to the
superconducting subsystem is also possible and has been recently studied in a
number of theoretical and experimental works \cite{Krivoruchko, Bergret_IPE,
Bergeret_IPE_Clean, Lofwander, Faure, Xia, Di_Bernardo, Lee_NatPhys, Salikhov,
Khaydukov, Nagy, Ovsyannikov}. This inverse proximity effect is related to the spin
polarization of electrons forming the Cooper pair near the S/F interface and
results in the small magnetization of the superconducting surface layer at the
depth of the order of the Cooper pair size, i.e., the superconducting
coherence length $\xi_{0}\sim1-10~nm$. Experimentally, it has been
observed with the help of muon spin rotation techniques \cite{Di_Bernardo,
Lee_NatPhys}, nuclear magnetic resonance \cite{Salikhov} and neutron scattering measurements \cite{Khaydukov, Nagy,
Ovsyannikov}.

It is commonly believed that neglecting this short range inverse proximity
phenomenon we can assume the magnetic field to remain completely trapped
inside the F film even when the electron transfer between the films is
possible. This conclusion is based on the obvious observation that an isolated
infinite single-domain ferromagnetic film with the in-plane magnetization does
not produce the stray magnetic field in the outside. In this paper we show
that, contrary to this belief, the direct proximity effect is always
responsible for the exciting the supercurrents flowing inside the ferromagnet
itself and, thus, for the appearance of the compensating Meissner
supercurrents in the superconducting subsystem. The appearance of these
currents is accompanied by the generation of the magnetic field in the S film
which decays at the distances of the order of the London penetration depth
$\lambda$ which can well exceed the coherence length in type-II
superconductors (see Fig.~\ref{Fig_System}). From the experimental point of
view, this electromagnetic proximity phenomenon reminds the Aharonov-Bohm
effect \cite{AB} since the current inside the attached superconductor is induced by the
ferromagnetic layer which  does not create the magnetic field outside the
layer in the absence of such superconducting environment. At the same time,
the true physical key point is that the wave function penetrating the
ferromagnet is responsible for this effect, when the vector-potential of the F
magnetization unavoidably generates current in quantum system described by a
common Cooper pair wave function (superconductor and a region near S/F
interface). Quite surprisingly this long-range electromagnetic (orbital)
contribution to the inverse proximity effect has been overlooked in all
previous studies devoted to this subject while the relevant magnetic fields
strongly exceed the ones induced by the spin polarization discussed in
\cite{Bergret_IPE, Bergeret_IPE_Clean} and should dominate in the experiments.
Indeed, in \cite{Lee_NatPhys, Khaydukov} the induced magnetic field was
observed in S/F hybrid systems at distances much larger than all relevant
superconducting coherence lengths. The existing theories of proximity effect
fail to provide the interpretation of these results, while our approach
provides a natural explanation of these phenomena. Thus, the current-carrying states appear to be unavoidable for the SF systems
and should be taken into account in the design of the devices of
superconducting spintronics \cite{Linder, Eschrig}.

\begin{figure*}[ptb]
\includegraphics[width=0.8\textwidth]{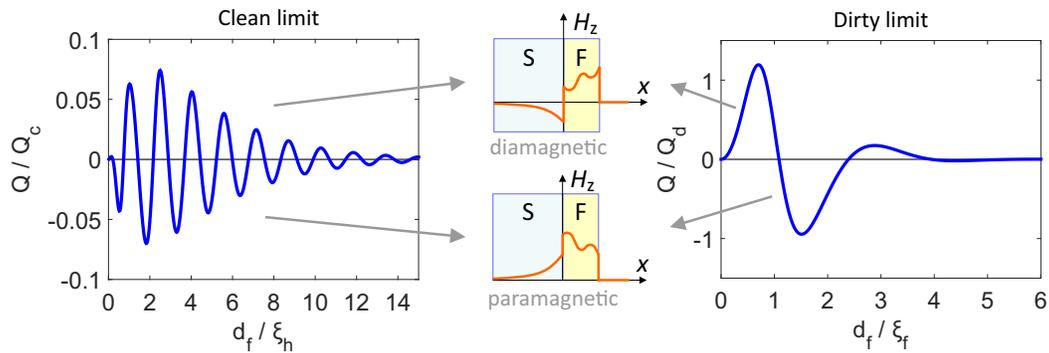}\caption{The dependence of the
magnetic kernel $Q$ on the ferromagnet thickness for the clean and dirty limits. For the clean limit we take $h=10\pi T$ and choose the temperature $T$ in a way that $\Delta=2\pi T$. Also we define the values $Q_d=\pi^2\sigma_f\xi_f^2\tanh(\Delta/2T)/(2\hbar c^2)$ and $Q_c=e^2\nu_0\xi_h^2(v_F/c)^2$ where $\sigma_f$ is the conductivity of the ferromagnet and $\nu_0$ is the density of states at the Fermi level per unit spin projection and per unit volume.}%
\label{Fig_Kernel}%
\end{figure*}

Further consideration illustrating the above qualitative arguments is
organized in two steps: (i) first, we consider a simple model assuming a
phenomenological form of the relation between the supercurrent $\mathbf{j}$
and vector potential $\mathbf{A}$ in conditions of the long range inverse
proximity effect; (ii) second, we present the results of microscopic
calculations of the $\mathbf{j}(\mathbf{A})$ relation which support and
justify the phenomenological findings. Accounting for the proximity effect we
write the Maxwell equation in the form:
\begin{equation}
\mathrm{rot}~\mathrm{rot}\mathbf{A}=\frac{4\pi}{c}\left(  \mathbf{j}%
_{s}+\mathbf{j}_{m}\right)  , \label{Eq_Maxwell}%
\end{equation}
where $\mathbf{j}_{s}$ is the Meissner current and $\mathbf{j}_{m}%
=c~\mathrm{rot}\mathbf{M}$ is the current associated with the magnetization
$\mathbf{M}$.

For the sake of definiteness we consider a bilayer consisting of the
superconductor of the thickness $d_{s}\gg\lambda$ and a ferromagnet of the
thickness $d_{f}\ll\lambda$ with the uniform magnetization $\mathbf{M}%
_{0}=M_{0}\mathbf{e}_{z}$. We choose the $x$-axis perpendicular to the layers
with $x=0$ at the S/F interface (see Fig.~\ref{Fig_System}). First, we assume
the local London relation which is relevant, e.g., for dirty S/F sandwiches:
\begin{equation}
\label{local}\mathbf{j}_{s}(x)=-\frac{c}{4\pi}\frac{1}{\lambda^{2}%
(x)}\mathbf{A}(x),
\end{equation}
where the London screening parameter $\lambda^{-2}$ becomes dependent on $x$.
The penetration of Cooper pairs into the F film gives rise to the supercurrent
there and also slightly modifies the Meissner response of the S layer in the
small region of the thickness of the order of the superconducting coherence
length $\xi$ from the S/F interface. The latter modification can even vanish
if the conductivity of the ferromagnet is much smaller than the normal
conductivity of the S layer. So, let us, first, neglect the changes of
$\lambda$ inside the superconductor and assume it to be equal to the constant
value $\lambda_{0}$. Then the solution of Eq.~(\ref{Eq_Maxwell}) in the S
layer reads $A_{y}(x)=A_{0}\exp(x/\lambda_{0})$ where $A_{0}$ is a constant
and inside the F layer $A_{y}(x)=A_{0}+4\pi M_{0} x$. Here we neglect the
spatial variations of the vector potential in the ferromagnet related to the
weak London screening since the typical scale of such variation is of the
order of $\lambda\gg d_{f}$. Thus, the exact solution of Eq.~(\ref{Eq_Maxwell}) is redundant for most realistic structure and material parameters and we restrict ourselves to the approximate calculation procedure described below.  To find $A_{0}$ we integrate
Eq.~(\ref{Eq_Maxwell}) over the width of the ferromagnet accounting the
relation $B_{z}=\partial A_{y}/\partial x$. The magnetic current
$\mathbf{j}_{m}=c~\mathrm{rot}\mathbf{M}$ integrated over the sample thickness
is zero and, thus, the relation between the magnetic field $B_{z}(d_{f})$
outside the sample and the field $B_{z}(0)=A_{0}/\lambda_{0}$ inside the
superconductor close to the S/F interface takes the form
\begin{equation}
\label{integr}B_{z}(d_{f})-B_{z}(0)=A_{0}\int_{0}^{d_{f}}\frac{dx^{\prime}%
}{\lambda^{2}(x^{\prime})}+4\pi M_{0}\int_{0}^{d_{f}}\frac{x^{\prime
}dx^{\prime}}{\lambda^{2}(x^{\prime})}.
\end{equation}
The first term in the r.h.s. of Eq.~(\ref{integr}) can be neglected since it
is of the order of $(d_{f}/\lambda)B_{z}(0)\ll B_{z}(0)$. Note that for the
same reason one can neglect the contribution coming from the renormalisation
of $\lambda^{-2}$ in the region of the width $\sim\xi$ inside the S layer if
the inverse proximity effect is not small. In the absence of the external
magnetic field $B_{z}(d_{f})=0$ and we finally obtain that the magnetic field
induced in the superconductor is
\begin{equation}
\label{Eq_result}B_{z}=-4\pi M_{0}Q\exp(x/\lambda_{0}),
\end{equation}
where $Q=\int_{0}^{d_{f}}\lambda^{-2}(x^{\prime})x^{\prime}dx^{\prime}$. Note
that in the conventional regime when the superconducting condensate
penetrating the F layer reveals a diamagnetic response with $Q>0$ the magnetic
field induced in the S layer is anti-parallel to the magnetization
$\mathbf{M}_{0}$. In the case when $d_{f}$ is of the order of the
superconducting coherence length in the ferromagnet $\xi_{f}$ the basic
estimate gives $Q\sim(\xi_{f}/\lambda)^{2}\sim10^{-2}$ for the S/F structures
based, e.g., on thin Nb films. Considering the experiments probing the change
in the magnetic moment in SF structures when cooling down through the
superconducting critical temperature and taking the typical magnetization
corresponding to Fe or Co films $4\pi M_{0}\sim10^{4}~\mathrm{Oe}$ we find
$B_{z}\sim10^{2}~\mathrm{Oe}$ which is an easily measurable value.

More accurate estimate for the value $Q$ can be obtained within the Usadel
formalism for dirty S/F bilayers \cite{Champel}. In Fig.~\ref{Fig_System}(c) we plot the typical profile of the screening parameter inside the F layer while the behavior of the kernel $Q$ is shown in the right panel of Fig.~\ref{Fig_Kernel}. 
Remarkably, the value $Q$ oscillates as a function of $d_{f}$ being either
positive or negative which corresponds to anti-parallel or parallel directions
of $\mathbf{B}$ and $\mathbf{M}_{0}$ for different values of $d_{f}$. 
Somewhat similar results for the magnetic response have been previously found for the S/F structures with a low transparent barrier at the interface in \cite{BVE}, but the magnetic field generated inside the superconductor was neglected.

\begin{figure}[b]
\includegraphics[width=0.35\textwidth]{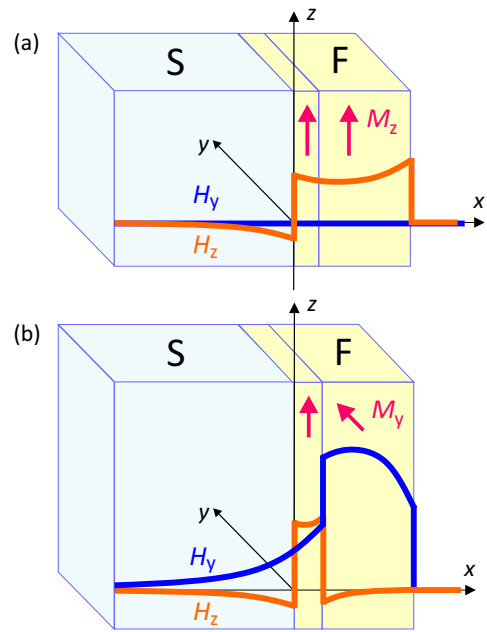}\caption{The profiles of the spontaneous magnetic field in the superconductor-ferromagnet-ferromagnet trilayers when the magnetic moments in the two F layer are (a) parallel to each other and (b) perpendicular to each other.}
\label{Fig_SFF}%
\end{figure}

Interestingly, the value $Q$ and, thus, the magnetic field in the S layer can
be substantially increased provided the magnetization in the F layer has a
non-collinear structure - this is relevant to the recent experiments
\cite{Lee_NatPhys}. To illustrate the origin of this effect let us assume that the ferromagnet
consists of two layers: one layer F$_{1}$ occupying the region $0<x<d_{1}$ has
the magnetization $M_{0}$ along the $z$-axis while another layer F$_{2}$ of
the thickness $d_{2}$ has the same magnetization $M_{0}$ but directed along
the $y$-axis. We choose $d_{2}$ to be much larger than the normal metal
coherence length $\xi_{n}=\sqrt{D_{f}/T}$ but still much less than $\lambda$.
The vector potential in the F bilayer has two components: $\mathbf{A}%
=\left(A_{0y}+4\pi M_{0}x\right)\mathbf{e}_{y}+ A_{0z}  \mathbf{e}_{z} \mathrm{~for~}0<x<d_{1}$ and $\mathbf{A}=\left(  A_{0y}+4\pi M_{0}d_1\right)  \mathbf{e}_{y}+\left[  A_{0z}-4\pi M_{0}(x-d_{1})\right]
\mathbf{e}_{z}\mathrm{~~for~}d_{1}<x<d_{1}+d_{2}$. The functions $A_{y0}(x)$
and $A_{z0}(x)$ vary over the distances $\sim\lambda$ and are almost constants
in the F film. Integrating Eq.~(\ref{Eq_Maxwell}) analogous to the case of the
S/F bilayer we find the magnetic field induced in the S layer:
\begin{equation}
B_{z}=-4\pi M_{0}Q_{z}e^{x/\lambda_{0}},~~~B_{y}=-4\pi M_{0}Q_{y}%
e^{x/\lambda_{0}}, \label{SFF_res}%
\end{equation}
where
\begin{equation}
Q_{z}=\int_{0}^{d_{1}}\frac{x^{\prime}dx^{\prime}}{\lambda^{2}(x^{\prime}%
)},~~Q_{y}=\int_{d_{1}}^{d_{1}+d_{2}}\frac{(x^{\prime}-d_{1})dx^{\prime}%
}{\lambda^{2}(x^{\prime})}. \label{Q12_def}%
\end{equation}
In Fig.~\ref{Fig_SFF} we schematically show the spatial profiles of the spontaneous magnetic field for parallel and perpendicular orientations of the magnetic moments in the ferromagnetic layers. Remarkably, for perpendicular orientation the estimate for the value $Q_y$ based on the Usadel formalism gives $Q_y\sim(\xi_n/\lambda)Q_z\gg Q_z$ where 
$\xi_n=\sqrt{D_f/T}\gg\xi_f$. As a result, in the
experiments where the angle $\theta$ between the magnetic moments in two F
layers can be tuned one should observe a strong increase in the magnetic
moment of the S layer for $\theta=\pi/2$ as compared to the case $\theta=0$ (see Fig.~\ref{Fig_SFF}).
Namely this behavior has been recently discovered in experiments with
the Au/Nb/ferromagnet structures \cite{Lee_NatPhys}.

Note that all described phenomena should become more pronounced provided one
deals with the clean S/F and S/F$_{1}$/F$_{2}$ sandwiches. In this case the
Cooper pairs penetrate over much larger distances into the ferromagnet which
strengthen the proximity effect and, thus, the amplitude of the induced
magnetic field in the superconductor. In the clean limit the relation
$\mathbf{j}_{s}(\mathbf{A})$ becomes non-local and we may write it in a very
generic form
\begin{equation}
\label{nonlocal}\mathbf{j}_{s}(x)=-\frac{c}{4\pi}\int\mathbf{A}(x^{\prime
})K(x,x^{\prime})dx^{\prime}.
\end{equation}
Deep inside the superconductor for $|x|\gg\xi_{N}=\hbar v_{F}/T$ the kernel
$K$ in this relation has the standard London form characterized by the
penetration depth $\lambda_{0}$. Then the profile of the magnetic field
generated in the superconductor is again determined by Eq.~(\ref{Eq_result})
but the value Q now can be expressed through the nonlocal kernel $K$ (see supplementary material):
\begin{equation}
\label{Q_phenom_res}Q=\int_{-x_{0}}^{d_{f}}dx\int_{0}^{d_{f}}dx^{\prime
}x^{\prime}\left[  K(x,x^{\prime})-\lambda_{0}^{-2}\delta(x-x^{\prime}%
)\theta(-x)\right]  ,
\end{equation}
where $\theta(x)$ is the Heaviside step function and the integration limit
$x_{0}$ satisfies the condition $\xi_{N}\ll x_{0}\ll\lambda_{0}$.

The calculations of the screening parameter $Q$ are presented in the supplementary material. In Fig.~\ref{Fig_Kernel}
we show the typical dependence of the value $Q$ on the ferromagnet thickness.
The interference between different quasiparticle trajectories results in the
oscillations of $Q(d_{f})$ with the period of the order of the length $\xi
_{h}=\hbar v_F/h$ which is known to characterize clean ferromagnets
\cite{Buzdin_RMP}. The envelope of the oscillating $Q$ increases as
$\propto(d_{f}/\xi_{N})^{2}$ at small $d_{f}/\xi_{N}$ values [here $\xi
_{N}=\hbar v_{F}/(4\pi T)$] and decays exponentially at $d_{f}\gg\xi_{N}$.

Up to now several experimental papers reported the evidences of the
long ranged (at distances larger than $\xi$) magnetic field generation in S/F
systems which could be naturally explained by the above theory. In
particular, in Ref.~\cite{Khaydukov} the polarized neutron reflectometry was
used to study the magnetic field profile in a single V(40~nm)/Fe(1~nm)
bilayer. Below $T_{c}$ these measurements indicate the appearance of the
spontaneous magnetic field penetrating the S layer at the distance
$\sim20~\mathrm{nm}$ from the S/F interface. This distance definitely exceeds
the coherence length $\xi_{0}\sim5\div10~\mathrm{nm}$ for vanadium. Another
manifestation of the long ranged field generation was recently reported in
\cite{Lee_NatPhys}, where the muon spin-rotation experiments in
Au/Nb/ferromagnet structure revealed a remote magnetic field in Au at the
large distance (more than 50 nm which strongly exceeds the $\xi$ value in Nb
films) from the ferromagnet. In \cite{Lee_NatPhys} the composite F layer was
used allowing a non-collinear magnetic configuration. The remote field was
much more pronounced in a perpendicular configuration in accordance with the
above arguments regarding the generation of the long-range magnetic moment in
SFF' systems. All these observations are pretty hard to explain by the
standard theory of the inverse proximity effect characterized by a rather
short length scale $\sim\xi_{0}$. In contrast, our results clearly demonstrate
the current and magnetic field generation at much larger distances
$\sim\lambda\gg\xi$ from the S/F interface. 
Note that repeating the experiments \cite{Lee_NatPhys} without the applied external field one may expect the generation of the magnetic field in the Au layer solely by the spontaneous current predicted in our work.

\begin{figure}[t!]
\includegraphics[width=0.45\textwidth]{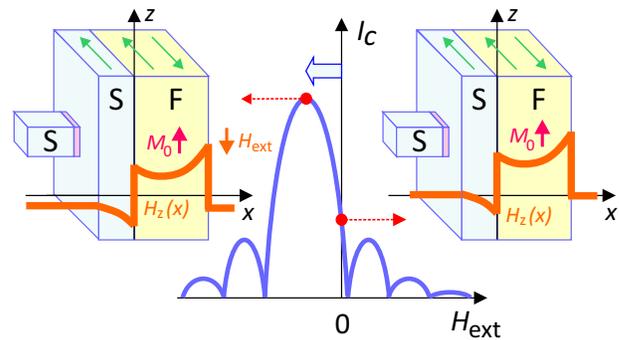}\caption{Shift in the Fraunhofer critical current oscillations for the Josephson junction with one electrode covered by the ferromagnetic layer. }%
\label{Fig_CritCurr}%
\end{figure}

Sure, the muon spin-rotation experiments permit the direct measurement of the spontaneous magnetic field in S/F structures. An alternative way to detect the currents associated with these spontaneous magnetic fields can be based on the use of different types of local transport probes  positioned at the outer boundary of the superconductor. Scanning, e.g., the outer surface by the normal metal tip of the scanning tunneling microscope one can measure the changes in the local density  of states caused by the Doppler shift of the quasiparticles energy in the presence of the superflow \cite{Berthod, Karapetrov, MaggioAprile}.  For the case of the superconducting probe one can also propose a simple and elegant experimental setup revealing this effect. It consists of the Josephson junction where one of the electrodes has the thickness of the order of $\lambda$ and is covered by the ferromagnetic layer (see Fig.~\ref{Fig_CritCurr}). Certainly, the superconducting probe should be small enough not to perturb the measured magnetic field distribution.  The electromagnetic proximity effect should result in the shift of the Fraunhofer dependence of the critical current on the external magnetic field. Note that the proper choice of the magnetic field can compensate this shift and restore the bare critical current (see Fig.~\ref{Fig_CritCurr}).

To sum up, we have revealed a very general mechanism of the long-range
electromagnetic proximity effect in the S/F structures which results in the
strong spread of the stray magnetic field into the superconductor from the
ferromagnet. The screening currents accompanying this magnetic field spread
appear in the region where in the normal state all the stray fields are
completely absent. The only nonzero electromagnetic characteristic in this
region is the vector potential which is usually an unobservable quantity. In
this sense the current generation which accompanies the superconducting
transition  illustrates the crucial role of the vector potential in quantum
physics of the SF structures.

\section*{Supplementary material}

See supplementary material for a microscopical
calculation of the electromagnetic response in the dirty and clean limits.

\section*{Acknowledgements}

The authors thank G. Karapetrov and Zh. Devizorova for valuable comments. This work was supported by the French ANR "SUPERTRONICS" and OPTOFLUXONICS, EU COST CA16218 Nanocohybri,  Russian Science Foundation under Grant No. 15-12-10020 (S.V.M.), Foundation for the advancement of theoretical physics ``BASIS'', Russian Presidential Scholarship SP-3938.2018.5 (S.V.M.) and Russian Foundation for Basic Research under Grant No. 18-02-00390 (A.S.M.). A.B. would like to thank the Leverhulme Trust for supporting his stay in Cambridge University.

\renewcommand{\theequation}{S\arabic{equation}}

\section*{Supplementary material for ``Electromagnetic proximity effect in planar superconductor-ferromagnet structures''}

\setcounter{equation}{0}

\subsection{Magnetic response: dirty limit}

To estimate the value $Q$ in the dirty limit we use  the Usadel
formalism. The spatial profile $\lambda^{-2}(x)$ is determined
by the anomalous Green function $\hat{f}(x)=(f_{s}+\mathbf{f}_{t}\hat{\sigma
})i\sigma_{y}$ which is a $2\times2$ matrix in the spin space ($\hat{\sigma}$
is the vector of the Pauli matrices):
\begin{equation}
\lambda^{-2}(x)=\frac{16\pi^{2}T\sigma}{\hbar c^{2}}\sum_{\omega>0}\left(
\left\vert f_{s}\right\vert ^{2}-\left\vert \mathbf{f}_{t}\right\vert
^{2}\right)  . \label{Lambda_Usadel}%
\end{equation}
In Eq.~(\ref{Lambda_Usadel}) the summation is performed over the positive
Matsubara frequencies $\omega$, $\sigma$ is the normal state conductivity and
inside the F layer the singlet and triplet components $f_{s}$ and
$\mathbf{f}_{t}$ satisfy the Usadel equations
\begin{equation}
D\partial_{x}^{2}f_{s}=2\omega f_{s}+2i\mathbf{h}\mathbf{f}_{t},~~~D\partial
_{x}^{2}\mathbf{f}_{t}=2\omega\mathbf{f}_{t}+2i\mathbf{h}f_{s}, \label{Usadel}%
\end{equation}
where $D$ is the diffusion coefficient and $\mathbf{h}$ is the exchange field.
For simplicity let us assume the conductivity of the F layer $\sigma_{f}$ to
be much smaller than the one of the superconductor ($\sigma_{s}$) which allows
to impose the rigid boundary conditions at the S/F interface for the $\hat{f}$
function: $f_{s}=f_{s0}$ and $\mathbf{f}_{t}=0$ where $\Delta$ is the gap
function and $f_{s0}=\Delta/\sqrt{\omega^{2}+\Delta^{2}}$. Assuming $\Delta$
to be real we can immediately write the solution of Eqs.~(\ref{Usadel}):
$f_{s}=\mathrm{Re}(F)$, $f_{tz}=i~\mathrm{Im}(F)$ where $F=f_{s0}\cosh\left[
q(x-d_{f})\right]  /\cosh\left(  qd_{f}\right)  $ and $q^{2}=2(\omega
+ih)/D_{f}$. The functions $f_{s}(x)$ and $f_{tz(x)}$ are oscillating and,
therefore, there are regions where $\left\vert f_{tz}\right\vert >\left\vert
f_{s}\right\vert $ and the local Meissner response is paramagnetic.

To calculate the value $Q$ for the S/F bilayer in the dirty limit it is convenient to rewrite the expression for the screening parameter $\lambda^{-2}$ through the function $F=f_s+f_{tz}$:
\begin{equation}\label{Lambda_Usadel}
\lambda^{-2}(x)=\frac{16\pi^2 T\sigma}{\hbar c^2}\sum_{\omega>0}{\rm Re}\left(F^2\right).
\end{equation}
Note that the frequencies $\omega$ enter only the amplitude $f_{s0}$ which allows to calculate the sum in Eq.~(\ref{Lambda_Usadel}) analytically:
\begin{equation}\label{sum}
\sum_{\omega>0}\frac{\Delta^2}{\omega^2+\Delta^2}=\frac{\Delta}{4T}\tanh\left(\frac{\Delta}{2T}\right).
\end{equation}
Also the value $Q$ is defined by the following integral: 
\begin{equation}\label{int}
\int\limits_0^{d_f}\frac{x\cosh^2\left[q(x-d_f)\right]dx}{\cosh^2\left(qd_f\right)}=\frac{q^2d_f^2-1}{4q^2\cosh^2\left(qd_f\right)}+\frac{1}{4q^2}.
\end{equation}
The last term in Eq.~(\ref{int}) is purely imaginary and it drops out from the value $Q$. Finally, after algebraic transformations we obtain:
\begin{equation}\label{Q_res}
Q=\frac{\pi^2 \sigma_f\xi_f^2\Delta}{2\hbar c^2}\tanh\left(\frac{\Delta}{2T}\right){\rm Im}\left[\frac{q^2d_f^2-1}{\cosh^2(qd_f)}\right],
\end{equation}
where $\xi_f=\sqrt{D_f/h}$. 


\subsection{Magnetic response: clean limit}

To calculate the magnetic response kernel in the clean limit we use the approach analogous to the one in [1, 2] which is based on the solution of the Eilenberger equation [3]. We consider a clean S/F bilayer with
$d_{f}\ll\lambda$ and $d_{s}\gg\lambda$. Inside the S layer the Eilenberger equations for the quasiclassical Green functions read
\begin{equation}
\label{Eq_Eilenberger_S}\hbar\mathbf{v}_{F}\hat{\mathbf{D}}f=-2\omega
f+2\Delta g,~\hbar\mathbf{v}_{F}\hat{\mathbf{D}}^{*}f^{\dag}=2\omega f^{\dag
}-2\Delta^{*} g,
\end{equation}
where $g$ is the normal Green function, $f$ and $f^{\dag}$ are the anomalous
Green functions satisfying the normalization condition $f^{\dag}f+g^{2}=1$,
$\hat{\mathbf{D}}=\left(  \nabla+2ie\mathbf{A}/\hbar c\right)  $ is the
gauge-invariant momentum operator, $\omega$ is the Matsubara frequency, and
$\mathbf{v}_{F}=v_{x}\hat{\mathbf{x}}_{0}+v_{y}\hat{\mathbf{y}}_{0}+v_{z}%
\hat{\mathbf{z}}_{0}$ is the vector of the quasiparticle velocity. For
simplicity we neglect the spatial variations of the gap potential $\Delta$
inside the S layer which is accurately justified in the vicinity of $T_{c}$ or
in the case of strong mismatch between Fermi velocities in the S and F layers
which damps the proximity effect. Note, however, that such variations of
$\Delta$ should not qualitatively change the final results, so our simple
model makes sense even at low temperatures. Inside the F layer we get:
\begin{equation}
\label{Eq_Eilenberger_F}\hbar\mathbf{v}_{F}\hat{\mathbf{D}}f=-2(\omega
+ih)f,~\hbar\mathbf{v}_{F}\hat{\mathbf{D}}^{*}f^{\dag}=2(\omega+ih)f^{\dag}.
\end{equation}
We focus only on the solutions of Eq. (\ref{Eq_Eilenberger_F}) near the S/F
interface in the region of the thickness $\sim(\hbar v_{F}/T)\ll\lambda$,
where the deviations of the electromagnetic response from the bulk London
relation $\mathbf{j}(\mathbf{A})$ are most pronounced. So, again, we put
$A_{y}(x)= A_{0} +4\pi M_{0} x$ neglecting the variations of $A_{y}(x)$ close
to the interface. 

The analytical solution of the Eilenberger equation for the function $g$ allows us to calculate the superconducting current flowing along S/F interface:
\begin{equation}\label{Eq_I_def}
j_y(x)=-4 \pi e\nu_0T\sum_{\omega>0}\left<{\rm Im}\left[g(x)\right] v_y\right>,
\end{equation}
where $\nu_0$ is the density of states at the Fermi level per unit spin projection and per unit volume, and the brackets denote averaging over the Fermi surface: $\left<...\right>=(4\pi)^{-1}\int_0^{\pi}d\theta\int_0^{2\pi}(...)\sin\theta d\varphi$. Here  $\theta$ and $\varphi$ are the spherical angles at the Fermi surface so that $v_x=v_F\cos\theta$ and $v_y=v_F\sin\theta\cos\varphi$. Further, it is convenient to represent this current as the sum $j_y(x)=j_M+j_{surf}(x)$, where $j_M=-(4/3) (e^2Tv_F^2\nu_0A_0(0)/\hbar c)\sum_{\omega>0}(\Delta^2/\Omega^3)$ coincides with the standard Meissner current flowing in the bulk superconductor and $j_{surf}(x)$ is the surface current induced by the magnetization. Finally, integrating $j_{surf}(x)$ over $x$ we find the desired value of the magnetic response kernel $Q$.


\bigskip 

\noindent [1] A. D. Zaikin and G. F. Zharkov, Zh. Eksp. Teor. Fiz. \textbf{81}, 1781 (1981).

\noindent [2] A. D. Zaikin, Sol. St. Commun. \textbf{41}, 533 (1982).

\noindent [3] N. B. Kopnin, in {\it Theory of Nonequilibrium Superconductivity.} The International Series of Monographs on Physics, Vol. 110 (Clarendon Oxford, 2001).


\begin{thebibliography}{0}

\bibitem{Buzdin_RMP} A.\ I.\ Buzdin, Rev. Mod. Phys., \textbf{77}, 935 (2005).

\bibitem{Nat1} A. Avsar, J. Y. Tan, T. Taychatanapat, J. Balakrishnan, G. K. W. Koon, Y. Yeo, J. Lahiri, A. Carvalho, A. S. Rodin, E. C. T. O'Farrell, G. Eda, A. H. Castro Neto, and B. \"{O}zyilmaz, Nat. Commun. \textbf{5}, 4875 (2014).

\bibitem{Nat2} T. Shoman, A. Takayama, T. Sato, S. Souma, T. Takahashi, T. Oguchi, K. Segawa, and Y. Ando, Nat. Commun. \textbf{6}, 6547 (2015).

\bibitem {Linder} J. Linder, J. W. A. Robinson, Nature Phys. \textbf{11}, 307 (2015).

\bibitem {Eschrig} M. Eschrig, Rep. Prog. Phys. \textbf{78}, 104501 (2015).

\bibitem{Bergeret_RMP} F.\ S.\ Bergeret, A.\ F.\ Volkov and K.\ B.\ Efetov, Rev. Mod. Phys. \textbf{77}, 1321 (2005).

\bibitem{Krivoruchko} V.\ N.\ Krivoruchko and E.\ A.\ Koshina, Phys. Rev. B \textbf{66}, 014521 (2002).

\bibitem{Bergret_IPE} F.\ S.\ Bergeret, A.\ F.\ Volkov, and K.\ B.\ Efetov, Phys. Rev. B \textbf{69}, 174504 (2004).

\bibitem{Bergeret_IPE_Clean} F.\ S.\ Bergeret, A.\ Levy Yeyati, and A.\ Mart\'{\i}n-Rodero, Phys. Rev. B \textbf{72}, 064524 (2005).

\bibitem{Lofwander} T. L\"{o}fwander, T. Champel, J. Durst, and M. Eschrig, Phys. Rev. Lett. \textbf{95}, 187003 (2005).

\bibitem{Faure} M.\ Faure, A.\ Buzdin and D.\ Gusakova, Physica C \textbf{454}, 61 (2007).

\bibitem{Xia} J.\ Xia, V.\ Shelukhin, M.\ Karpovski, A.\ Kapitulnik, and A.\ Palevski, Phys. Rev. Lett. \textbf{102}, 087004 (2009).

\bibitem{Di_Bernardo} A. Di Bernardo, Z. Salman, X. L. Wang, M. Amado, M. Egilmez, M. G. Flokstra, A. Suter, S. L. Lee, J. H. Zhao, T. Prokscha, E. Morenzoni, M. G. Blamire, J. Linder, and J. W. A. Robinson, Phys. Rev. X \textbf{5}, 041021 (2015).

\bibitem{Lee_NatPhys} M.\ G.\ Flokstra, N.\ Satchell, J.\ Kim, G.\ Burnell, P.\ J.\ Curran, S.\ J.\ Bending, J.\ F.\ K.\ Cooper, C.\ J.\ Kinane, S.\ Langridge, A.\ Isidori, N.\ Pugach, M.\ Eschrig, H.\ Luetkens, A.\ Suter, T.\ Prokscha, and S. L. Lee, Nat. Phys. \textbf{12}, 57 (2016).

\bibitem{Salikhov} R. I. Salikhov, I. A. Garifullin, N. N. Garif'yanov, L. R. Tagirov, K. Theis-Br\"{o}hl, K. Westerholt, and H. Zabel, Phys. Rev. Lett. \textbf{102}, 087003 (2009).

\bibitem{Khaydukov} Yu.\ N.\ Khaydukov, B.\ Nagy, J.-H.\ Kim, T.\ Keller, A.\ R\"{u}hm, Yu.\ V.\ Nikitenko, K.\ N.\ Zhernenkov, J.\ Stahn, L.\ F.\ Kiss, A.\ Csik, L.\ Botty\'{a}n, V.\ L.\ Aksenov, Zh. Eksp. Teor. Fiz. \textbf{98}, 107 (2013).

\bibitem{Nagy} B. Nagy, Yu. Khaydukov, D. Efremov, A. S. Vasenko, L. Mustafa, J.-H. Kim, T. Keller, K. Zhernenkov, A. Devishvili, R. Steitz, B. Keimer and L. Bottyr\'{a}n, Europhys. Lett. \textbf{116}, 17005 (2016).

\bibitem{Ovsyannikov} G. A. Ovsyannikov, V. V. Demidov, Yu. N. Khaydukov, L. Mustafa, K. Y. Constantinian, A. V. Kalabukhov, D. Winkler, JETP \textbf{122}, 738 (2016).

\bibitem{AB} Y. Aharonov and D. Bohm, Phys. Rev. \textbf{115}, 485 (1959).

	
\bibitem{Champel} T. Champel, M. Eschrig, Phys. Rev. B \textbf{72}, 054523 (2005).


\bibitem {BVE} F. S. Bergeret, A. F. Volkov, K. B. Efetov,  Phys. Rev. B \textbf{64}, 134506 (2001).


\bibitem{Berthod} C. Berthod, Phys. Rev. B \textbf{88}, 134515 (2013).

\bibitem{Karapetrov} S. A. Moore, G. Plummer, J. Fedor, J. E. Pearson, V. Novosad, G. Karapetrov, and M. Iavarone, Appl. Phys. Lett. \textbf{108}, 042601 (2016).

\bibitem{MaggioAprile} C. Berthod, I. Maggio-Aprile, J. Bru\'{e}r, A. Erb, and C. Renner, Phys. Rev. Lett. \textbf{119}, 237001 (2017).

\end{thebibliography}
\end{document}